\documentclass{icya}
\begin{document}

\title[Comparison between particle and fluid approximations to dust dynamics]{Comparison between particle and fluid approximations to dust dynamics}

\author[Joanna Dr\c{a}żkowska]{Joanna Dr\c{a}żkowska\thanks{Toruń Centre for Astronomy, Nicolaus Copernicus University, Toruń, Poland}, Michał Hanasz\thanksmark{1}, Kacper Kowalik\thanksmark{1}}


\newcommand{\DTP}{{\sffamily\bfseries Desktop Publishing}}
\newcommand{\memoir}{{\sffamily\bfseries memoir}\ }
\newcommand{\texlive}{\TeX live\ }
\newcommand{\bibtex}{Bib\TeX\ }

\iabstract{%
We present a new particle module of the magnetohydrodynamic (MHD) Piernik code. The original multi-fluid grid code based on the Relaxing Total Variation Diminishing (RTVD) scheme has been extended by addition of dust described within the particle approximation. The dust is now described as a system of interacting particles. The particles can interact with gas, which is described as a fluid. In this poster we introduce the scheme used to solve equations of motion for the particles and present the first results coming from the module. The results of test problems are also compared with the results coming from fluid simulations made with Piernik-MHD code. The comparison shows the most important differences between fluid and particle approximations used to describe dynamical evolution of dust under astrophysical conditions.
}

\maketitle

\section{Piernik-MHD}  
Piernik is a multi-fluid grid MHD code based on the RTVD conservative scheme by \citet{Jin95} and \citet{tracpen}. Piernik can be used to examine dynamics of ionized or neutral gas, as well as dust treated as a pressureless fluid. The code computes conservative fluid variables (fluid density, momentum, total energy density) for each cell of the grid. The basic scheme has been extended by addition of many facilities which are useful in astrophysical fluid-dynamical simulations, e.g. shearing-box boundary conditions, Ohmic resistivity module and selfgravity module.
See \citet{pier1,pier2,pier3,pier4} for more details.

\section{Particle module}
Dust can be described in fluid and particle approximations. 
In the particle module of PIERNIK-MHD code the dust component is described as a system of independent particles that can interact with each other. The particles can also interact with gas considered as a fluid. For each particle, equation of motion is solved using the scheme described in the next subsection.

\subsection{Scheme}
To solve the equation of motion for dust particles we use the scheme known as Verlet leap-frog method. In this algorithm, the velocities are calculated at time $t+\frac{1}{2}dt$ and used to calculate the positions, $r$, at time $t+dt$. In this way, the velocities leap over the positions, then the positions leap over the velocities.

Generally, the scheme can be noted as:
    \begin{equation}
      r(t+dt)=r(t)+v(t+\frac{1}{2}dt)dt,
    \end{equation}
    \begin{equation}
      v(t+\frac{1}{2}dt)=v(t-\frac{1}{2}dt)+a(t)dt.
    \end{equation}

\section{Results}
To compare fluid and particle approximations applied for the dust component we carried out several test problem simulations with the same initial conditions applied in both approaches.

\begin{figure}[!ht]
\centering
\includegraphics[width=.45\textwidth]{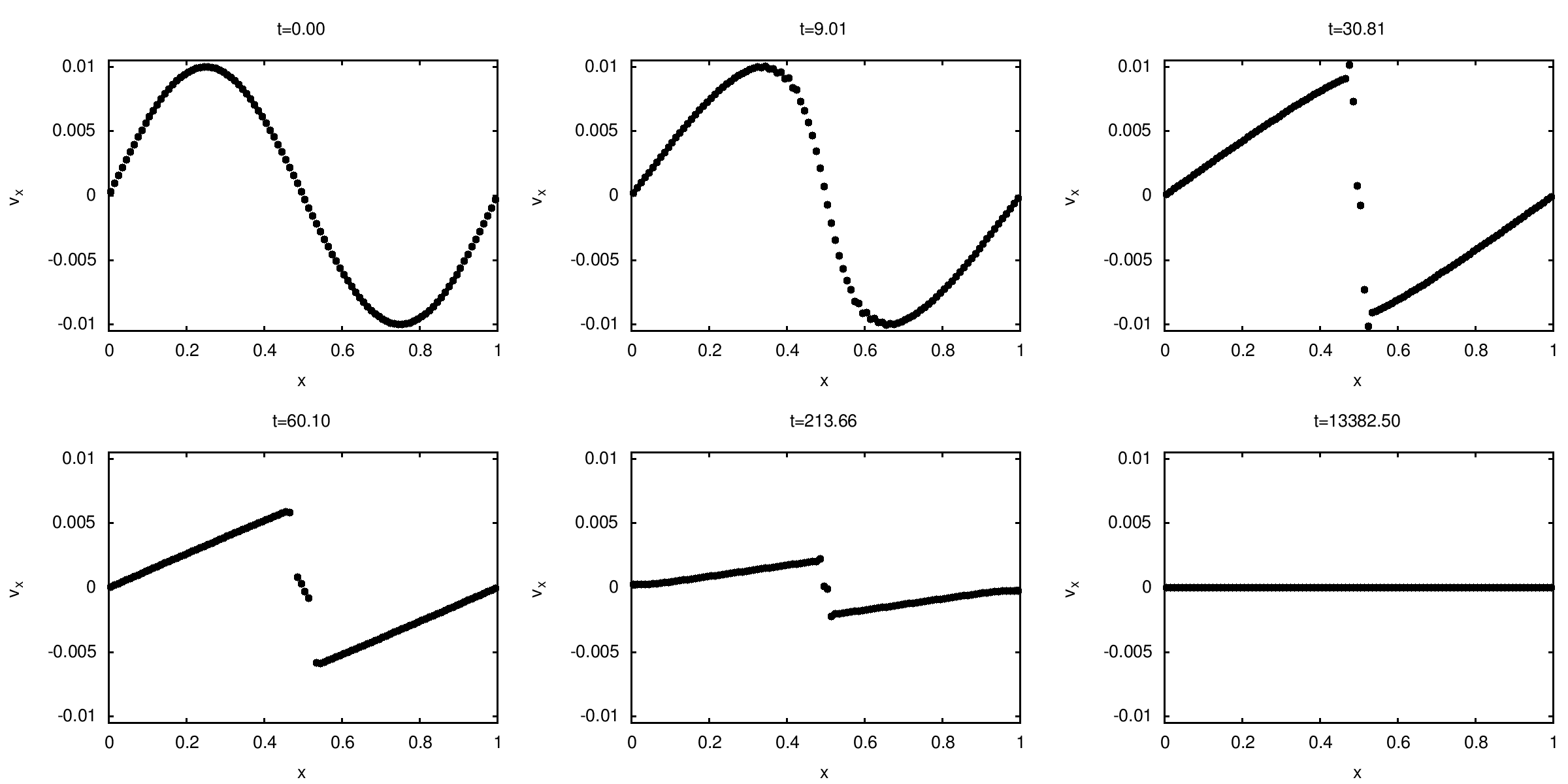}
\caption[]{Sinusoidal wave simulation result coming from the fluid approximation}
\label{fig:piersin}
\end{figure}

The first test problem relies on the analysis of 1D sinusoidal velocity perturbation. The fluid approximation result, which can be veryfied by an analytical solution of the Burger's equation \citep{toro}, displays a conversion of the initial sinusoidal velocity profile into the sawtooth profile and then smoothing until the flat profile (figure~\ref{fig:piersin}). The discontinuity in the velocity profile can be interpreted as shock front.

\begin{figure}[!ht]
\centering
\includegraphics[width=.45\textwidth]{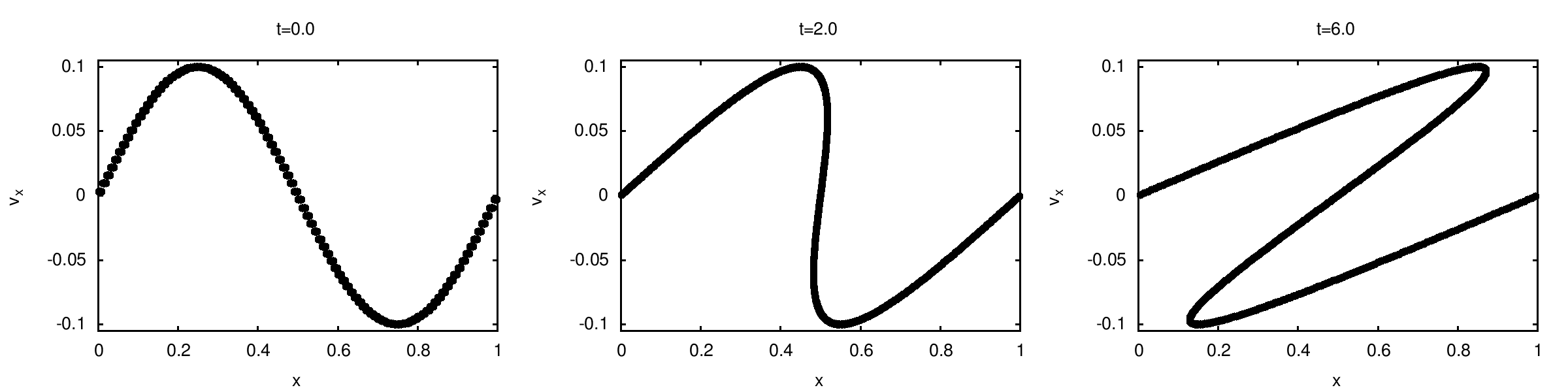}
\caption[]{Sinusoidal wave simulation result coming from the particle model in case of noninteracting particles}
\label{fig:noint}
\end{figure}

In the case of noninteracting particles the particle model leads to multiple velocity values in the velocity profile (figure~\ref{fig:noint}). To avoid the unphysical evolution of the particle system we have introduced interaction between particles. The interaction is analogous to inelastic collisions. The particles stick when they meet each other in the same grid cell.

\begin{figure}[!ht]
\centering
\includegraphics[width=.45\textwidth]{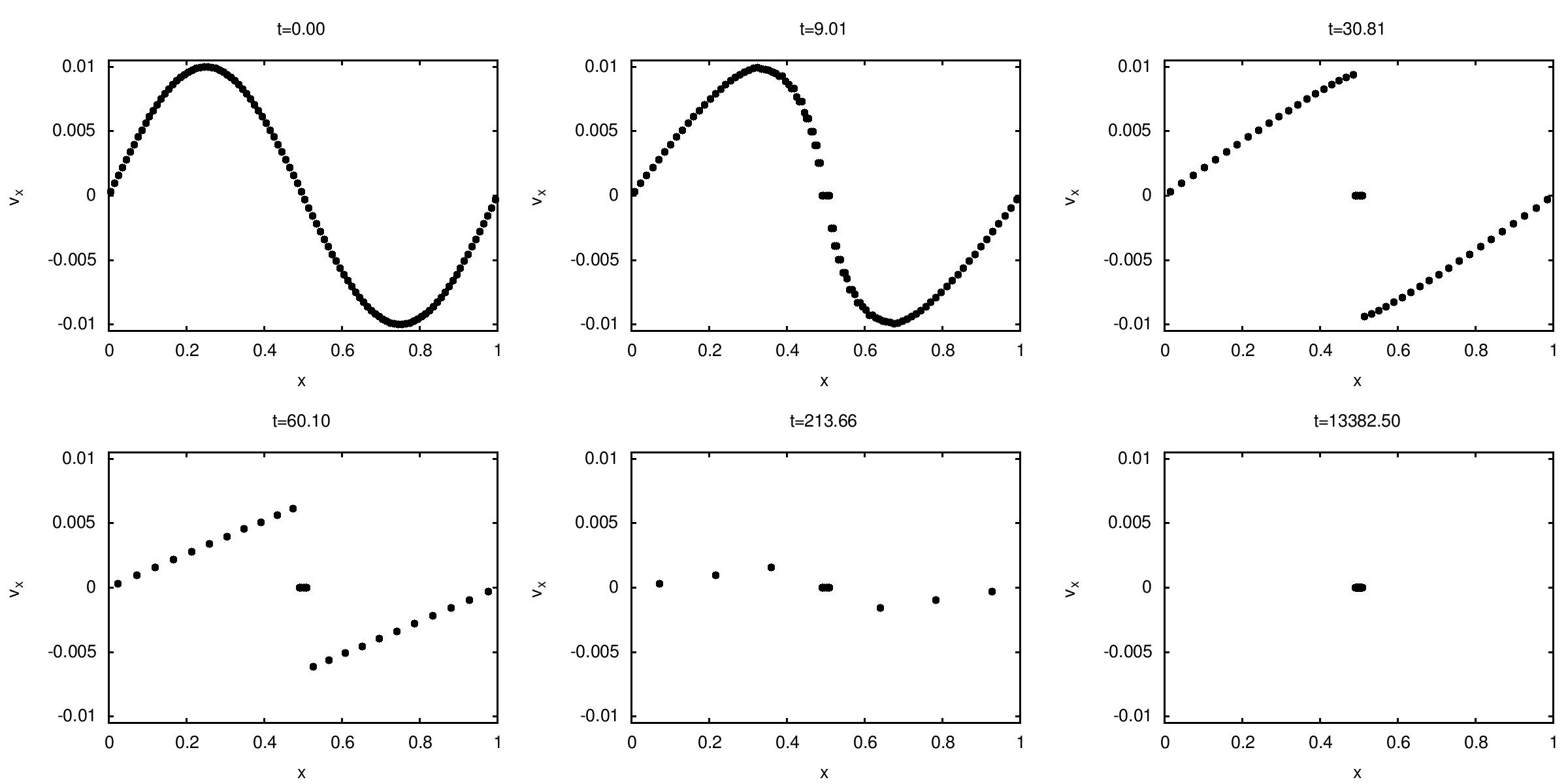}
\caption[]{Sinusoidal wave simulation result coming from the particle model}
\label{fig:modsin}
\end{figure}

The result coming from the particle approach with interactions taken into account (figure \ref{fig:modsin}) appears different than the result given by the fluid approximation, because the particles group together into clusters.
In the fluid simulations all the physical quantities are computed for every cell of the domain, even if density is very small. In the particle simulations the values of physical quantities are specified only in the particles locations. The fluid density profile at the end of the fluid simulation is represented by one peak of density. Respectively, at the end of the particle simulation all the particles are grouped together into one aggregate.

\begin{figure}[!ht]
\centering
\includegraphics[width=.45\textwidth]{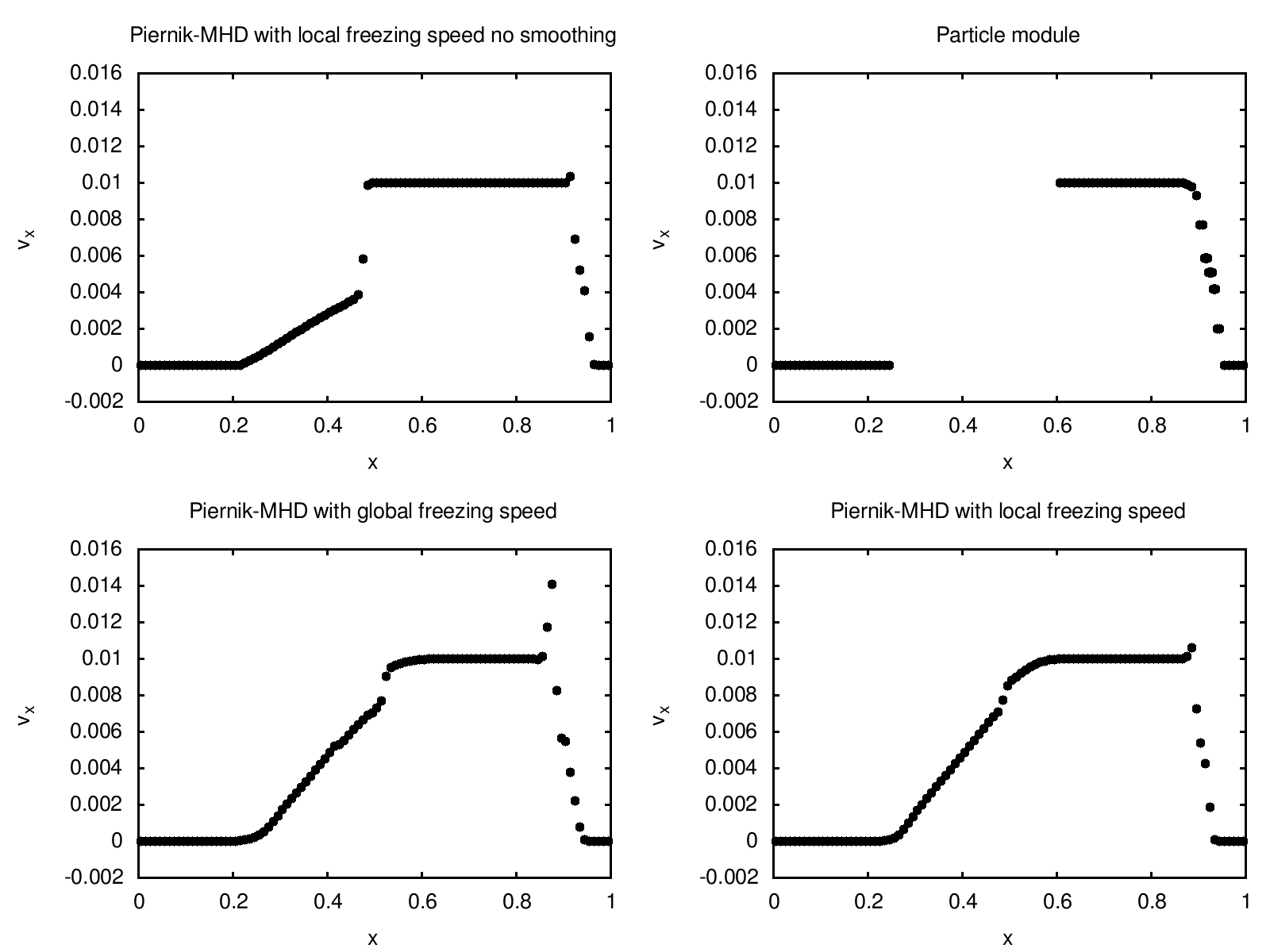}
\caption[]{Comparison of square wave simulations results}
\label{fig:square}
\end{figure}

Another comparison test of the two methods is related to the evolution of the 1D square velocity perturbation. The results are displayed in the figure \ref{fig:square}. 
We note that the borders of rarefying shock front area are different in the both metodhs. The profile given by the fluid simulation is not consistent with the analytical solution of the Burger's equation.
It turnes out that the difference in the results is a numerical artefact related to discontinuities in the first derivative of the so called freezing speed function (see  \citet{tracpen}.
Freezing speed is a quantity specific for the RTVD method, used  to decompose vectors of conservative variables and their fluxes into left-moving and right-moving waves.
Freezing speed can be computed locally as velocity for each cell or globally as the maximum velocity in the whole domain.
When we introduced the local freezing speed smoothing (or just used the global freezing speed), the dust velocity profile became significantly more similar to the velocity profile obtained in the particle approximation (the lower plots in figure \ref{fig:square}).

\begin{figure}[!ht]
\centering
\includegraphics[width=.4\textwidth]{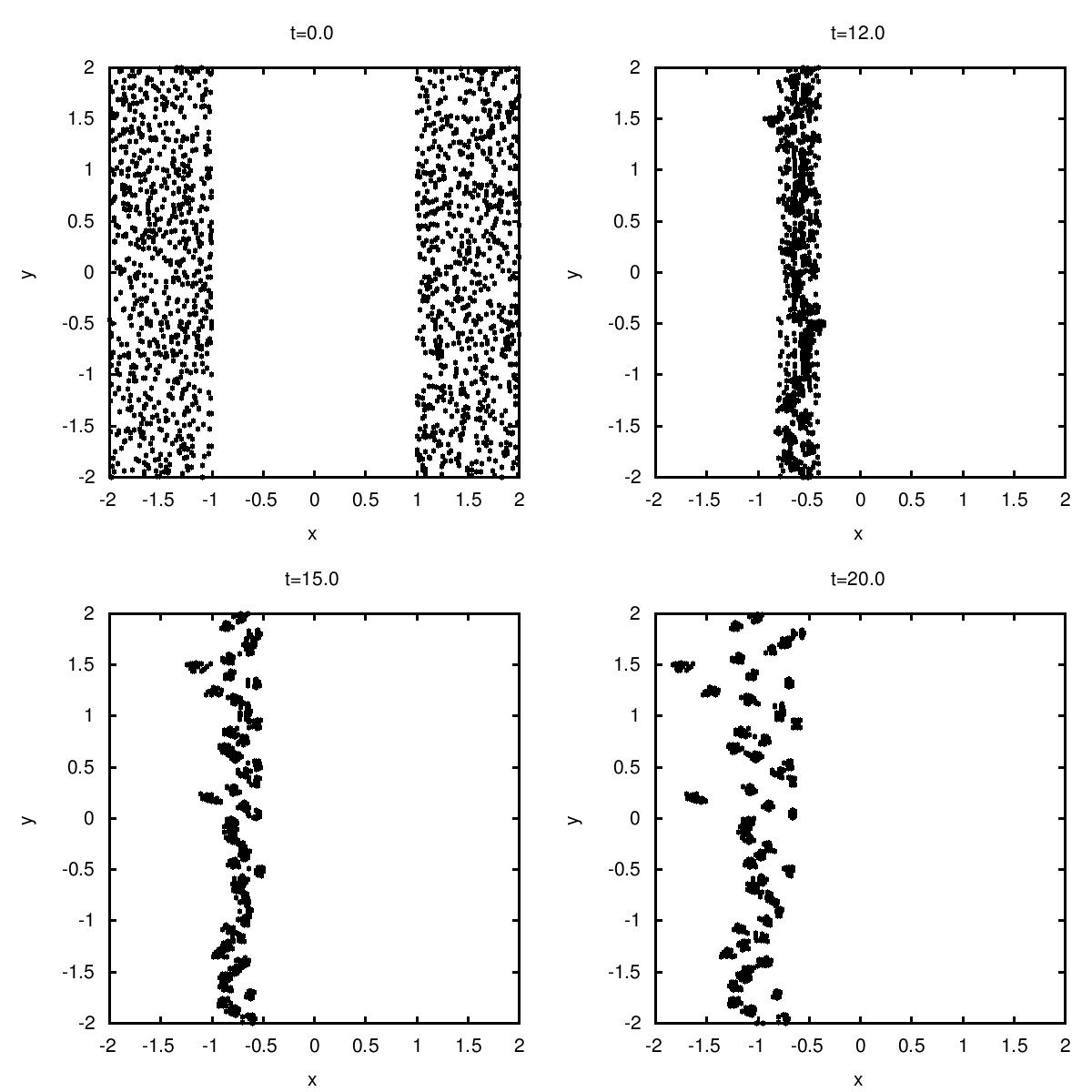}
\caption[]{Dust fronts collision simulation result coming from the particle module}
\label{fig:frontsmod}
\end{figure}

The differences between the fluid and particle approaches are much more significant in 2D tests. We present results of two identical dust fronts collision test. The left front moves initially two times faster than the right one. The fronts fragmentation (figure \ref{fig:frontsmod}) is an effect of the particle simulation. The fragmentation is probably caused by the finite grid resolution and the fluctuations of the mean momentum in the cells containing a small number of particles.

\begin{figure}[!ht]
\centering
\includegraphics[width=.4\textwidth]{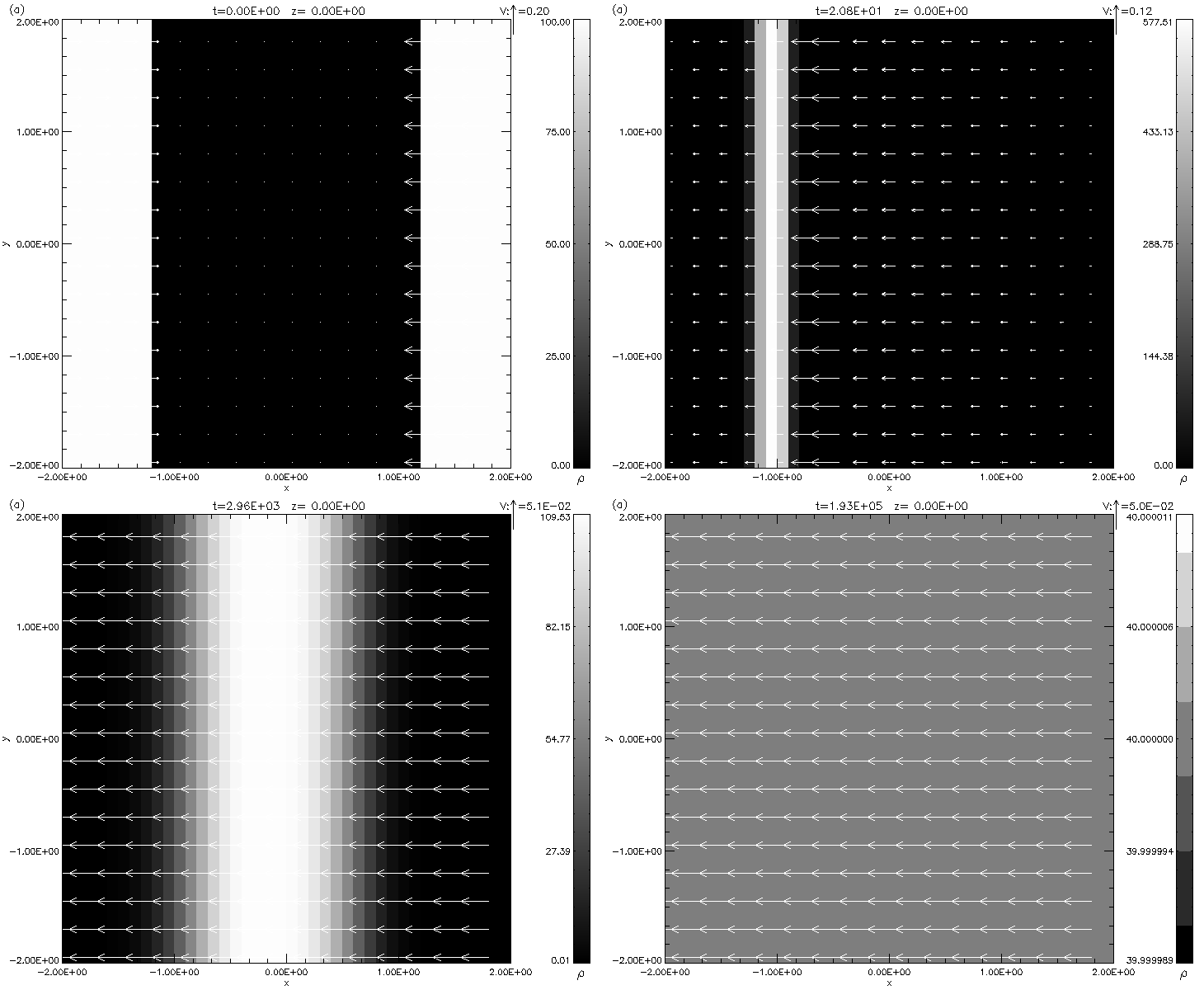}
\caption[]{Dust fronts collision simulation result coming from the fluid approximation}
\label{fig:frontspier}
\end{figure}

In the fluid simulation  (figure \ref{fig:frontspier}) the fronts merge and move together with the velocity resulting from the momentum conservation law. On a longer timescale the front diffuses over the whole computational domain.

\acknowledgments{This work was partially supported by by Polish Ministry of Science and Higher Education through the grants 92/N-ASTROSIM/2008/0, and by Nicolaus Copernicus University through the grant No. 365--A.}



\begin{thebibliography}{7}
\providecommand{\natexlab}[1]{#1}
\providecommand{\url}[1]{\texttt{#1}}
\expandafter\ifx\csname urlstyle\endcsname\relax
  \providecommand{\doi}[1]{doi: #1}\else
  \providecommand{\doi}{doi: \begingroup \urlstyle{rm}\Url}\fi

\bibitem[{Hanasz} et~al.(2008{\natexlab{a}}){Hanasz}, {Kowalik}, {Wóltański},
  \&~{Pawłaszek}]{pier1}{Hanasz}, M., {Kowalik}, K., {Wóltański}, D.,
  \&~{Pawłaszek}, R.
\newblock In {K.~Go\'zdziewski}, , editor, \textsl{Extra-solar planets in
  multi-body systems}, December 2008{\natexlab{a}}, arXiv:0812.2161.
\bibitem[{Hanasz} et~al.(2008{\natexlab{b}}){Hanasz}, {Kowalik}, {Wóltański},
  \&~{Pawłaszek}]{pier3}{Hanasz}, M., {Kowalik}, K., {Wóltański}, D.,
  \&~{Pawłaszek}, R.
\newblock In {M.~de~Avillez}, , editor, \textsl{The Role of Disk-Halo
  Interaction in Galaxy Evolution: Outflow vs. Infall?}, December
  2008{\natexlab{b}}, arXiv:0812.2799.
\bibitem[{Hanasz} et~al.(2008{\natexlab{c}}){Hanasz}, {Kowalik}, {Wóltański},
  {Pawłaszek}, \&~{Kornet}]{pier2}{Hanasz}, M., {Kowalik}, K., {Wóltański},
  D., {Pawłaszek}, R., \&~{Kornet}, K.
\newblock In {K.~Go\'zdziewski}, , editor, \textsl{Extra-solar planets in
  multi-body systems}, December 2008{\natexlab{c}}, arXiv:0812.4839.
\bibitem[{Hanasz} et~al.(2009){Hanasz}, {Kowalik}, {Wóltański},
  \&~{Pawłaszek}]{pier4}{Hanasz}, M., {Kowalik}, K., {Wóltański}, D.,
  \&~{Pawłaszek}, R.
\newblock In {M.~de~Avillez}, , editor, \textsl{The Role of Disk-Halo
  Interaction in Galaxy Evolution: Outflow vs. Infall?}, January 2009, arXiv:0901.0104.
\bibitem[Jin \&~Xin(1995)]{Jin95}Jin, S. \&~Xin, Z., 1995, Comm. Pure Appl.
  Math., \textbf{48}, 235--276.
\bibitem[{Toro}(1999)]{toro}{Toro}, E.~F., 1999, \textsl{{Riemann Solvers and
  Numerical Methods for Fluid Dynamics: a practical introduction}}.
\newblock Springer.
\bibitem[{Trac} \&~{Pen}(2003)]{tracpen}{Trac}, H. \&~{Pen}, U.-L., March 2003,
  PASP, \textbf{115}, 303--321, arXiv:astro-ph/0210611
\end{thebibliography}
\end{document}